\documentclass[twocolumn,twoside]{revtex4}
\usepackage{graphicx}
\usepackage{amsmath,amssymb}
\usepackage{fancyhdr}
\newcommand{\nn}{\nonumber}
\newcommand{\eqs}[1]{\begin{align} #1 \end{align}}
\newcommand{\Lcal}[0]{\mathcal{L}}
\newcommand{\Ocal}[0]{\mathcal{O}}
\newcommand{\ovl}{\overline}
\newcommand{\diff}[3]{
\if 1#1  \frac{\mathrm{d} #2 }{\mathrm{d} #3 }
\else  \frac{\mathrm{d}^{#1} #2 }{\mathrm{d}#3^{#1} } \fi
}

\begin{document}

\title{Renormalization Effects on Electric Dipole Moments in Electroweakly Interacting Massive Particle Models}

\author{Takumi Kuwahara}
\affiliation{Center for Theoretical Physics of the Universe, Institute for Basic Science (IBS), Daejeon, 34126, Korea}

\begin{abstract}
The extended models of the standard model with a single Majorana fermion could be realized as the low-energy effective theory of the well-motivated ultraviolet models.
We study the electric dipole moments generated by the effective operators which are composed of the Majorana fermion and the standard model Higgs bosons, especially focusing on the renormalization effects of the effective operators.
We give the one-loop anomalous dimension of the effective operators from the scale where the operators are generated to the electroweak scale.
In this proceedings, we focused on the electric dipole moments from electroweak triplet and 5-plet.
We found the renormalization effects could give an enhancement factor being of the order of $\Ocal(10)$\% for a triplet model and being more than two in a 5-plet fermion model.
\end{abstract}

\maketitle

\thispagestyle{fancy}

\section{Introduction}

$CP$ violation of the standard model (SM) particles is a good probe of the models beyond the standard model (BSMs).
The electron electric dipole moment (EDM) in the SM is generated only at four-loop diagrams with inserted the CKM phase: atomic/molecular EDM is generated by both the electron EDM of the order of $10^{-44}~ e\,\mathrm{cm}$ and the $CP$-odd nucleon-electron four-Fermi interactions of the order of $10^{-38}~ e\,\mathrm{cm}$ \cite{Pospelov:2013sca} in the SM.
These predictions in the SM are much smaller than the updated bound on the electron EDM reported by ACME collaboration: $|d_e| < 1.1 \times 10^{-29}~e\,\text{cm} ~ (90\% \, \text{CL})$ \cite{Andreev:2018ayy}.
It would also be expected that the ACME experiment will improve the limit on electron EDM by almost two orders of magnitude at most in the future.
Since there should be generally new sources of $CP$ violation in the BSMs, the electron EDM could be sensitive to the new $CP$-violating sources.

The electroweakly interacting massive particle (EWIMP), which is charged under $SU(2)_L \times U(1)_Y$, could be a good candidate of the dark matter (DM).
Indeed, a quintuplet fermion is automatically stable due to the absence of decay operators up to dimension five \cite{Cirelli:2005uq,Cirelli:2007xd,Cirelli:2009uv}.
Moreover, EWIMP fermions could appear in low-energy effective models of specific ultraviolet (UV) completions.
For instance, triplet and/or doublet fermions could be the DM particles in split supersymmetric (SUSY) models \cite{ArkaniHamed:2004fb,Giudice:2004tc,ArkaniHamed:2004yi,Wells:2004di,Ibe:2006de,Hall:2011jd,Arvanitaki:2012ps,ArkaniHamed:2012gw,Ibe:2012hu}.
Therefore, the EWIMP models have been intensively studied in the context of collider physics and astrophysics.

We focus on the $CP$ violating interactions between EWIMPs and the SM Higgs, which arise from the dimension-five effective operator and lead to the electron EDM, with particular attention to the renormalization group (RG) effects.
Once we find the electron EDM in the future, it could be important to estimate the energy scale where new particles appear.
In particular, in order to determine the energy scale, the RG effects get more important when the UV scale is much higher than the mass scale of the EWIMP.

A part of the work \cite{Kuramoto:2019yvj} is presented in this proceeding.
Due to limitations of space, this proceeding has focused only on the electron EDM from Majorana EWIMP fermions.
We have also discussed the case of the Dirac EWIMP fermion, the nucleon EDMs, and EDMs in specific models in Ref.~\cite{Kuramoto:2019yvj}.

\section{EDMs in EWIMP Models \label{sec:EDM}}

Let $\chi$ be an $SU(2)_L$ $n$-plet Majorana EWIMP fermion.
This fermion contains an electromagnetically neutral component, and it could be a candidate of the DM.
Once we assume the neutral part dominantly composes the whole DM, the mass of the EWIMP is completely determined by the thermal abundance of the DM.
However, we do not assume that in this study, and then the EWIMP mass is still a free parameter.
The Lagrangian for $\chi$ is given by
\eqs{
\Lcal_\chi & = \frac12 \left(i \ovl \chi \gamma^\mu D_\mu \chi - M \ovl \chi \chi\right)
+ \dfrac{1}{2} \tilde C_s H^\dag H \ovl \chi^C i\gamma_5 \chi \, ,
\label{eq:CPvioint_M}
}
where $D_\mu$ is a gauge covariant derivative, and $M$ is the mass of the EWIMP fermion.
The last term denotes the dimension-five $CP$-odd interaction to the SM Higgs $H$.
A $CP$-even counterpart of the interaction could appear, but the $CP$-even operator is  irrelevant for our work.
Once we specify the UV completion, we can determine the effective coupling $\tilde C_s$ as a function of fundamental parameters, such as couplings and mass parameters, in the UV completion.
In this work, we treat $\tilde C_s$ as a free parameter.
In particular, since this operator could arise as a tree-level contribution, we define $\tilde C_s \equiv \xi/M_\mathrm{phys}$ where $M_\mathrm{phys}$ is the mass scale of a new particle appeared in the UV completion, and $\xi$ collectively denotes dimensionless couplings and $CP$ violation in the EWIMP sector.

Another dimension-five operator should appear in the Dirac EWIMP models, which is with insertion of the $SU(2)_L$ generators.
However, this operator vanishes in Majorana EWIMP models due to the real property of the EWIMP fermion.

\begin{figure}[h]
\centering
\includegraphics[width=45mm]{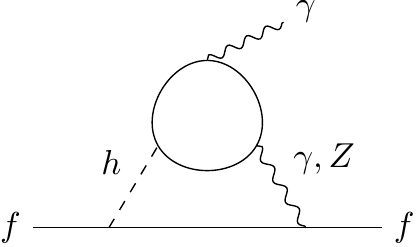}
\caption{
	The Barr-Zee diagrams inducing the electron EDM.
	The EWIMP multiplet runs in the inner loop.
} \label{fig:bzgraph}
\end{figure}

This effective coupling gives the electron EDM as the electroweak (EW) scale at the two-loop level as shown in Fig.~\ref{fig:bzgraph}.
The electron EDM is computed as follows \cite{Hisano:2014kua}.
\eqs{
d_e & = \frac{e \alpha_e m_e Q_e}{6 (4\pi)^3}
\frac{n(n^2 - 1)}{M} \tilde C_s f_0\left( \frac{M^2}{m_h^2} \right) \,,
}
where $m_h$ is the mass of the SM Higgs, $\alpha_e$ is the electromagnetic fine structure constant, and $f_0(x)$ is the loop function.
In general, as shown in Fig.~\ref{fig:bzgraph}, there is the $Z$-boson contribution as well for a light fermion $f$, but in the case of the electron EDM, the $Z$-boson contribution is accidentally suppressed.
One may wonder if there would be the $W$-boson contribution.
In the Majorana EWIMP fermion case, the $W$-boson contribution vanishes due to the cancellation between the contributions from the positive-charged fermions and the negative-charged fermions.
A complete expression of the EDM of a light fermion $f$, for the both case of the Majorana and Dirac EWIMPs, is given in Refs.~\cite{Hisano:2014kua,Nagata:2014aoa,Kuramoto:2019yvj}.

We should comment on the impact from $CP$-odd dimension-six effective operators.
The dimension-six operators are in principle suppressed by $M_\mathrm{phys}^2$, and there must also be a loop factor.
Thus, the contributions from the dimension-six operators could be subdominant for the case of $M_\mathrm{phys} \gg M \,, m_Z$.
When the high-energy input scale $M_\mathrm{phys}$ is close to $M$ and the EW scale, these contributions are sizable and the following RG effects would be less important.

\section{Renormalization Group Equation \label{sec:rge}}

The effective operator in Eq.~(\ref{eq:CPvioint_M}) could arise at the energy scale of $M_\mathrm{phys}$, while the electron EDM is computed at the EW scale.
The electron EDM can set severe constraint on the effective operator, so that  $M_\mathrm{phys}$ could be much larger than the EW scale.
When the hierarchy between $M_\mathrm{phys}$ and the EW scale is quite large, the RG evolution of the effective coupling is important to carry out precise calculations of EDMs.

\begin{figure}[h]
\centering
\includegraphics[width=80mm]{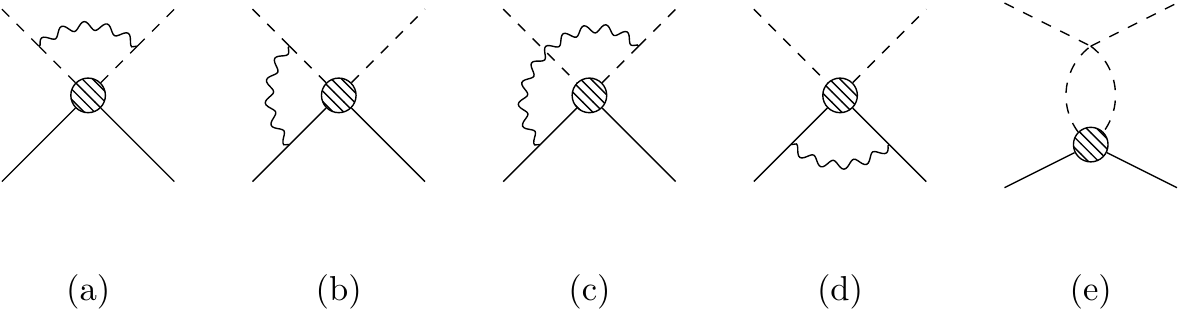}
\caption{One-loop vertex corrections to the effective operator (denoted by blobs).} \label{fig:oneloop}
\end{figure}

In this work, we compute the evolution of the effective coupling at the one-loop order.
The renormalization group equation (RGE) of the effective coupling among the Majorana EWIMP and the SM Higgs is defined by
\eqs{
\diff{1}{\tilde C_s(\mu)}{\ln \mu}
& = \gamma_s\tilde C_s(\mu) \,,
\label{eq:RGE}
}
with the renormalization scale $\mu$ and the anomalous dimension $\gamma_s$.
There are two contributions to the anomalous dimension: one is the wavefunction renormalization of the external Higgs legs and the external EWIMP legs, the other is the vertex correction.
In particular, dominant contributions
The vertex corrections are shown in Fig.~\ref{fig:oneloop}: the corrections (a)-(d) arise from the gauge interactions while and the correction (e) comes from the Higgs quartic interaction, defined by
\eqs{
V(H) = \frac{\lambda}{4} (H^\dag H)^2 \,.
}
Of course, there are diagrams where the inner propagator is attached to the external legs on the other sides of those in the correction (b)-(c).

We derive the one-loop anomalous dimension $\gamma_s$ for the $CP$-violating dimension-five operators in single EWIMP fermion extension of the SM
The anomalous dimension is obtained as follows.
\eqs{
\gamma_s = &  - \frac{1}{(4 \pi)^2} \left[6 g^2 \left( C_2(H) + C_2(\chi) \right) + 6 g^{\prime 2} Y_H^2 \right. \nn \\
& \left. - 3 \lambda - 6 y_t^2\right]
\label{eq:RGE_Wilson}
}
where $C_2$ is the quadratic Casimir invariant for $SU(2)_L$ representations: $C_2(H)$ for the representation of the SM Higgs, $C_2(\chi)$ for that of the EWIMP fermion, and $C_2(G)$ for an adjoint representation.
$g'$ is the $U(1)_Y$ gauge coupling, and $Y_H = 1/2$ is the hypercharge of the SM Higgs.
Since $C_2(\chi) = \frac14 (n^2 - 1)$ for an $SU(2)_L$ $n$-plet $\chi$, the effective coupling receives a large correction in the case of the large $n$ representation $\chi$.
Furthermore, when $n$ is so large that the Higgs quartic coupling and the top Yukawa coupling are negligible, the anomalous dimension is negative, and then the RG evolution makes the effective coupling enhanced at the low-energy scale.

\section{Results}

In this section, we discuss the numerical results of the electron EDM induced by the $CP$-odd effective operator.
The following fermions are considered in this study: a 5-plet fermion and a triplet fermion.
The former fermion is motivated by so called the minimal dark matter model, and the latter often appears in several UV completions, such as split supersymmetry models.
As we mentioned in Sec.~\ref{sec:EDM}, we do not assume that their neutral components can explain the whole DM abundance.

\begin{figure}[h]
\centering
\includegraphics[width=80mm]{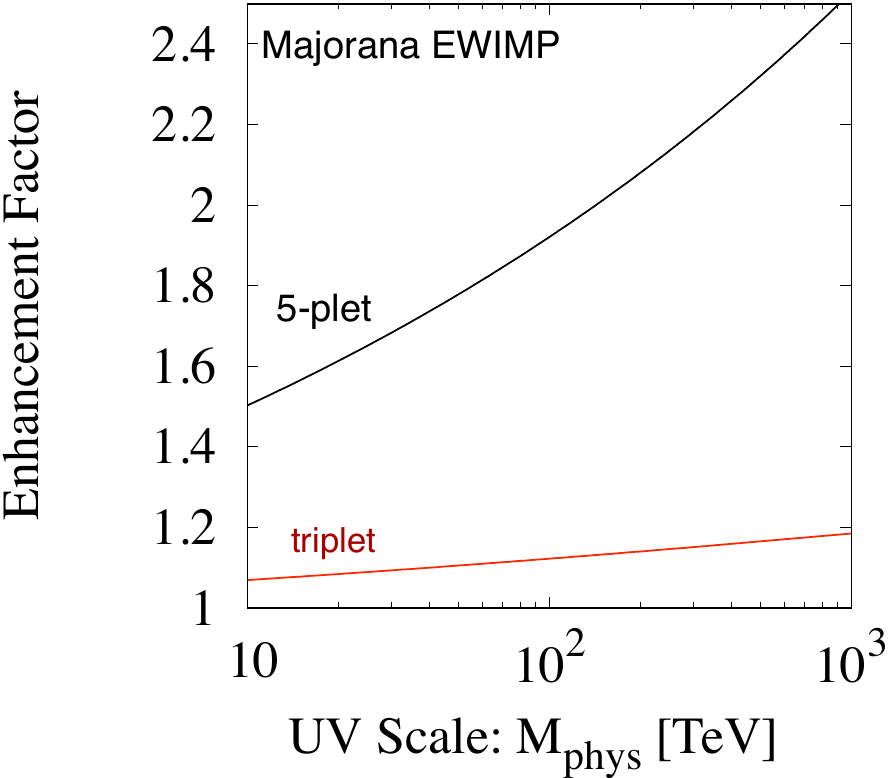}
\caption{
Enhancement factors in the Majorana EWIMP fermion models.
The red (black) line shows the enhancement factor in the triplet (5-plet) EWIMP model
} \label{fig:efactor}
\end{figure}

First, we show the impact of the RG effect on the electron EDM.
We define the enhancement factor as the ratio of the effective coupling $\tilde C_s(\mu)$ at the EW scale ($\mu = m_Z$) and the input scale ($\mu = M_\mathrm{phys}$) which the effective operator is generated.
\eqs{
(\text{Enhancement Factor}) \equiv \frac{\tilde C_s(m_Z)}{\tilde C_s(M_\mathrm{phys})} \,.
}
In order to solve Eq.~(\ref{eq:RGE}), the other dimensionless couplings should also follow the RGEs for them.
We use two-loop beta functions for gauge couplings and one-loop beta functions for the top Yukawa coupling and the Higgs quartic coupling in this study.

Fig.~\ref{fig:efactor} shows the enhancement factor in the Majorana EWIMP models.
This factor does not depend on the dimensionless parameter $\xi$ because we take the ratio of the effective coupling.
The computation of the electron EDM from the effective operator is performed at the EW scale, and thus there is also no dependence of the EWIMP mass $M$.

For the 5-plet Majorana fermion case, the enhancement factor is in the range of 1.6 to 2.4 for $M_\mathrm{phys} = 10~\mathrm{TeV} \, \text{-} \, 10^3~\mathrm{TeV}$, and then it could reach to a factor of two.
On the other hand, the electron EDM from the triplet EWIMP gets the enhancement factor of twenty percent at most.

\begin{figure}[h]
\centering
\includegraphics[width=80mm]{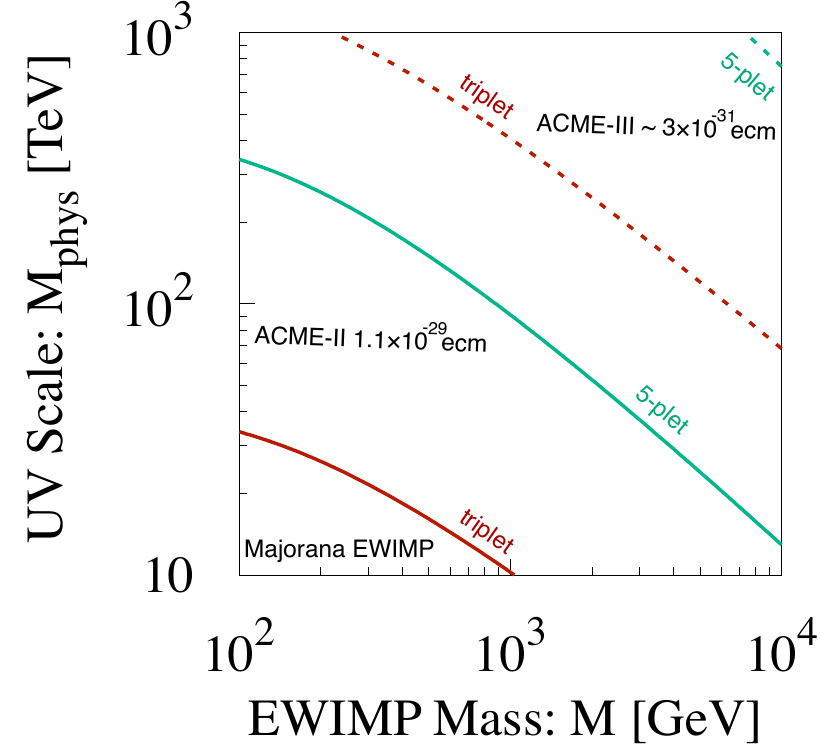}
\caption{
	Electron EDM in the Majorana EWIMP models.
	The current bound $|d_e| < 1.1 \times 10^{-29}~e\,\mathrm{cm}$ corresponds to solid lines, while the future projected sensitivity $|d_e| \sim 3 \times 10^{-31}~e\,\mathrm{cm}$ corresponds to dashed lines.
} \label{fig:eedm}
\end{figure}

Lastly, Fig.~\ref{fig:eedm} show the $M$-$M_\mathrm{phys}$ contour plot of the electron EDM from the Majorana EWIMP models.
The effective coupling is generated at the energy scale of $M_\mathrm{phys}$, and we assume the size of the dimensionless parameter as $\xi = 0.1$ in this figure.
We also assume that there is no other $CP$ violation in the SM effective field theory, so that the $CP$-odd dimension-five operator dominates the electron EDM.
The solid lines show the current bound on the electron EDM which is reported by ACME-II, and thus the regions below the solid lines is excluded.
The dashed lines show projected sensitivity at ACME-III.
The color difference indicates the difference of representations of EWIMP fermions: a triplet fermion (red) and a 5-plet fermion (green).

Without the RG effect, the EDMs are proportional to $M^{-1} M_\mathrm{phys}^{-1}$, so that contours in the $M$-$M_\mathrm{phys}$ contour plot should be straight.
Due to the presence of the RG effect, the EDM is enhanced in large $M_\mathrm{phys}$ region.
Fig.~\ref{fig:eedm} illustrates the enhancement in large $M_\mathrm{phys}$ region; the lines curve in that region.

The electron EDM from $SU(2)_L$ $n$-plet is proportional to $n(n^2-1)$, and thus the exclusion limit gets large for the large $n$-plet when the value $\xi$ is fixed.
The future sensitivity of the ACME-III, $|d_e| \sim 3 \times 10^{-31}~e\,\mathrm{cm}$, could reach to the whole parameter region for the case of the 5-plet Majorana fermion.

\section{Concluding Remarks}

The neutral components of electroweak multiplets could be a good candidate of the DM.
The searches for the EWIMPs are carried out extensively in collider physics and astrophysics.
In this proceeding, we focus on the sensitivity of EWIMPs in the precision frontier.
The fermionic EWIMP could have interactions to the SM Higgs via $CP$-odd dimension-five operators, and the operators lead the electron EDM.

To provide the theoretical calculation of the EDM with high accuracy, we have derived the anomalous dimensions for the dimension-five operators.
Actually, the current EDM experiments have a good sensitivity to the operators even if the cutoff scale is about 100~TeV, so that the RG effect can be significant.
In this proceeding, we discuss the RG effects in the Majorana EWIMP fermion models.
The RG effects give the enhancement factor of the order of $\Ocal(10)$~\% for the triplet EWIMP case and that of the order of $\Ocal(100)$~\% for the 5-plet EWIMP case.

In this proceeding, we do not discuss the nucleon EDMs from the dimension-five operators.
For the quark EDMs below the electroweak scale, we can use the leading order QCD correction provided for the $b \to s \gamma$ process \cite{Ciuchini:1993fk} (see also, Ref.~\cite{Degrassi:2005zd}).
Our work provides the same order corrections to the quark EDMs from the electroweak gauge, top Yukawa, and Higgs quartic couplings, so we completed the leading order corrections to the EDMs from the $CP$-odd dimension-five operators.

\begin{acknowledgments}
	I am grateful to Ryo Nagai and Wataru Kuramoto for fruitful collaboration.
	I would also like to thank the organizers of Flavor Physics and CP Violation Conference, 2019.
	This work was supported by IBS under the project code, IBS-R018-D1.
\end{acknowledgments}

\bigskip
\bibliography{ref}

\end{document}